\documentclass[prd,twocolumn,showpacs]{revtex4}%
\usepackage{graphicx}
\usepackage{amsmath}
\usepackage{amsfonts}
\usepackage{amssymb}
\usepackage{color}
\usepackage{amscd}
\usepackage{bm}

\begin{document}
\title{Space-time uncertainty relation from quantum and gravitational principles}
\author{Yi-Xin Chen}
\email{yxchen@zimp.zju.edu.cn}
\author{Yong Xiao}
\email{yongxiao2008@live.cn}

\affiliation{Zhejiang Institute of Modern Physics}
\affiliation{Zhejiang University, Hangzhou 310027, China}

\begin{abstract}
By collecting both quantum and gravitational principles, a
space-time uncertainty relation $(\delta t)(\delta
r)^{3}\geqslant\pi r^{2}l_{p}^{2}$ is derived. It can be used to
facilitate the discussion of several profound questions, such as
computational capacity and thermodynamic properties of the universe
and the origin of holographic dark energy. The universality and
validity of the proposed relation are illustrated via these
examples.
\end{abstract}
 \maketitle
\section{Introduction\label{s1}}
Any physical system in the quantum world is undergoing quantum
fluctuations, which introduce space and time uncertainties to the
system. Interestingly, general relativity (GR) conspired with
quantum mechanics (QM) have determined the severity of these
space-time uncertainties.

We firstly start from a theoretical question: What is the maximal
spatial resolution when using photons uniformly spread in a box of
size $l$ for observation. Statistical mechanics tells that the
photon gas has the energy $E\sim
\frac{k_{B}^{4}}{c^{3}h^{3}}l^{3}T^{4}$. To insure that the system
will not collapse to form a black hole, there must be a GR
limitation that $E\sim\frac{k_{B}^{4}
}{c^{3}h^{3}}l^{3}T^{4}\leqslant E_{bh}\sim\frac{c^{4}}{G}l$, where
$E_{bh}$ is the energy of a black hole of the same size. This
limitation leads to
$k_{B}T\leqslant\left(\frac{c^{7}h^{3}}{G}\right) ^{1/4}l^{-1/2}$.
The corresponding thermal wavelength is thus $\lambda
=\frac{hc}{k_{B}T}\geqslant l^{1/2}\left( \frac{hG}{c^{3}}\right)
^{1/4}=l^{1/2}l_{p}^{1/2}$, where
$l_{p}\equiv\left(\frac{hG}{c^{3}}\right)^{1/2}$ is the Planck
length. It gives the uncertainties in the positions of the photons
themselves, so one can not detect the system by virtue of these
photons with more precision than their shortest wavelength
$l^{1/2}l_{p}^{1/2}$. Such a limitation is firstly given by 't Hooft
in \cite{thooft}. Since the size of a physical system should be
larger than $l_{p}$, there is $\delta l\geqslant
l^{1/2}l_{p}^{1/2}\geqslant l_{p}$. It implies that when limited to
a smaller region, one can always employ higher energy photons and
detect finer structures until the Planck scale.

The corresponding resolution of the system with $\delta l\geqslant
l^{1/2}l_{p}^{1/2}$ reaches $\frac{l^{3}}{\left( \delta l\right)
^{3}}\sim A^{3/4}$, where $A$ is the boundary area of the system. To
find a holographic resolution \cite{susskind,bousso} which is
proportional to the area, namely $\frac{l^{3}}{\left(
 \delta l\right) ^{3}}\sim A$, we have to require $\delta l\geqslant
l^{1/3}l_{p}^{2/3}$. The latter uncertainty is consistent with
holographic principle and related to the unknown details of black
hole physics \cite{jack1}. And it is different from the
uncertainties of radiation systems which are described by local
quantum field theory (LQFT).

Notice that both the two uncertainty relation $\delta l\geqslant
l^{1/2}l_{p}^{1/2}$ and $\delta l\geqslant l^{1/3}l_{p}^{2/3}$ have
spread in the literature with various derivations and applications
\cite{thooft,baez,cohen,us,michael,jack1,jack2}. Firstly the two
relations can be written with a form of ultraviolet-infrared (UV-IR)
relations. Cohen et al \cite{cohen} have proposed that the UV-IR
relation for LQFT systems and for holographic systems are
respectively $l^{3}\Lambda ^{4}\leqslant l$ and $l^{3}\Lambda
^{3}\leqslant l^{2}$, or written as $\Lambda \leqslant l^{-1/2}$ and
$\Lambda \leqslant l^{-1/3}$. (Hereafter we shall set $G,\hbar
,c,k_{B}=1$ and not write $l_{p}$ or $m_{p}$ explicitly.) Since the
UV cutoff $\Lambda$ determines the minimal detectable lengths, that
is $\Lambda=\left(\delta l\right)^{-1}$, they directly correspond to
the two uncertainty relations above. Again, the different UV-IR
relations play an important role in the strict verification of the
entropy gap between LQFT and holographic systems, from $A^{3/4}$ to
$A$ \footnote{The entropy bound $A^{3/4}$ can be obtained by a
strict counting of non-gravitational collapse quantum states where
the UV-IR relation naturally presents itself \cite{us}. The entropy
gap from $A^{3/4}$ to $A$ also has its cosmological counterpart
\cite{barrow}, which brings about a huge numerical differences from
$10^{90}$ to $10^{120}$ at the present era of our universe. The
existence of such an entropy gap will be further verified in Section
\ref{s3.2}.}.

On the other hand, the two type of uncertainties could be understood
as the uncertainties in the measurement of a distance $l$. Realizing
such distance measurements through Gedanken experiments using clocks
and light signals, by carefully evaluating the quantum and
gravitational corrections to the measurement procedure, both $\delta
l\geqslant l^{1/2}$ and $\delta l\geqslant l^{1/3}$ can be derived
consistently \footnote{Note here it refers to the uncertainty in
which a distance $l$ can be measured. Such an uncertainty is
different from that in the measurement of a position located with a
macroscopic object. The latter is undoubtedly the minimal length of
nature, i.e. the Planck length $l_ {p}$ \cite{hsu1}, while the
former can be derived as a cumulative effect of $\pm l_{p}$ over the
distant $l$ \cite{michael}.} \cite{michael}. Obviously, which
$\delta l$ should be applied depends on which tools one use for
observation or transmitting signals: conventional photons or some
unknown holographic \textquotedblleft particles\textquotedblright.

Moreover, Ng has treated these uncertainties from a viewpoint called
quantum foam \cite{jack1,jack2}. That is treating the space-time
geometry as undergoing quantum fluctuations which manifest
themselves by the accuracy with which one can measure a distance
$l$, generally written as $\delta l\geqslant
l^{\alpha}l_{p}^{1-\alpha}$ with $\alpha=\frac{1}{2}$ and
$\alpha=\frac{1}{3}$ as its special cases.

The two type of quantum fluctuations also find applications in
cosmology to establish dark energy models. It has been found that
both of them can yield an energy density of the form
$\rho_{\Lambda}\sim l^{-2}$. Choosing the IR cutoff $l$ to be a
proper cosmological scale such as the size of cosmological horizons,
the energy density of this type defines the so-called
\textquotedblleft holographic dark energy (HDE)\textquotedblright\
models \cite{limiao,nojiri,hsu,m2,caironggen1,wei}.

Generally speaking, different systems are subjected to different
quantum fluctuations, up to what they are composed of. The
characteristic quantum fluctuation $\delta l \geqslant l^{1/2}$ is
applicable to LQFT systems with entropy bound $A^{3/4}$, while
$\delta l \geqslant l^{1/3}$ is applicable to strongly gravitational
systems like black holes with holographic entropy. Meanwhile,
$\delta l$ not only characterizes the quantum fluctuations and then
the space-time uncertainties within the interior of a physical
system, but also characterizes the measurement error to the size $l$
of the entire system.

In this Letter, we devote ourselves to gain more insights into the
significance of the quantum uncertainties rooted in a system. In
Section \ref{s2}, from a quantum computational perspective,
combining principles from GR and QM, we find a space-time
uncertainty relation $\left( \delta t\right)\left( \delta
r\right)^{3}\geqslant \pi r^{2}$, with $\delta t$ and $\delta r$
representing the severity of space-time fluctuations of the
constituents of the system at small scales. We find the relation
could be very useful in the cases where the analysis of fundamental
degrees of freedom plays an essential role. Several examples of this
kind will be included in Section \ref{s3}. The main physical
characteristics of them are extracted directly through the
space-time uncertainty relation and compared to the known results in
the literature.

\section{Space-time Uncertainty Relation\label{s2}}
A physical system could always be reduced to independent degrees of
freedom doing Boolean calculations. The shift of states on these
self-governed quantum bits leads to the evolvement of the entire
system. In this section, we shall start from such a quantum
computational perspective and derive a space-time uncertainty
relation.

Without loss of generality, we consider a globular computer of
radius $r$. Assume the computer is made up of $\frac{r^{3}}{(\delta
r)^{3}}=r^{3}\Lambda^{3}$ independent functional units which are
employed to store information and execute instructions, with $\delta
r\equiv\Lambda^{-1}$ representing the size of a single functional
unit. Every unit has an energy $\varepsilon$ for executing
operations. GR requires that the computer as a whole cannot has an
energy exceeding the mass of a black hole of the same size. Thus
\begin{align}
\varepsilon r^{3}\Lambda^{3}\leqslant E_{bh}= r/2.  \label{1}
\end{align}
For any independent degree of freedom or a quantum bit, the
Margolus-Levitin theorem \cite{levitin,seth1,seth2,seth3} determines
the minimal time it takes to finish an operation (a shift of states)
is $\delta t=\frac{\pi}{2\varepsilon}$, where $\varepsilon$ is the
energy distributed in this degree of freedom for it to execute
Boolean calculations. Together with Eq.\eqref{1}, we find a
space-time uncertainty relation
\begin{align}
(\delta t)(\delta r)^{3}\geqslant\pi r^{2}.  \label{2}
\end{align}
The above derivation to Eq.\eqref{2} is a heuristic one. Actually
the simplified picture above is closely related to realistic
physical systems. Taking the photon gas for example, the photons
within have an uncertainty in position which is $\delta r\geqslant
r^{1/2}$, as we have shown at the beginning of this Letter. In
addition, none of these $A^{3/4}$ uncorrelated photons (without
wave-packet overlapped) is fixed and permanently unchangeable.
Actually every of them is described by a quantum state evolving with
time according to the QM laws, specifically Margolus-Levitin theorem
here. The time interval it takes for a quantum bit or an independent
photon here evolving from one state to its orthogonal states is
typically $\delta t\geqslant r^{1/2}$. Obviously, it characterizes
the uncertainty or randomness of the quantum states of an isolated
photon in time direction. $\delta r$ and $\delta t$ together conform
to Eq.\eqref{2}. They characterize the quantum uncertainties of the
constituents of the system within the space-time, but not directly
the fluctuations of the space-time itself, unless one has introduced
some microscopic mechanism of quantum gravity and then deals with a
bulk of gravitons.

Whatever, it should be emphasized that Eq.\eqref{2} is not a
space-time uncertainty relation of usual type. Though the physical
system is composed of a large amount of fundamental units or degrees
of freedom, in determining their uncertainties within space-time, GR
has to be employed to limit the energy of the system as a whole.
Thus Eq.\eqref{2} can only be applied to the cases taking a global
viewpoint of measurement on the system, that is simultaneously
taking all of its fundamental units into consideration. Furthermore,
the deduced uncertainties $\delta r$ and $\delta t$ are relevant to
the size $l$ of the entire system. Actually it means that the
combination of GR an QM principles would inevitably cause an unusual
correlation between the global and local properties of a system like
the UV-IR relations \cite{cohen}. For general discussions about the
limit on space-time measurements, one may refer to
\cite{michael,hsu1} and references therein.

For LQFT systems like the conventional matter and radiation, one
usually require the space and time are treated equally
\cite{jack1,jack2}. Combining with the relation \eqref{2}, the
uncertainties of $\delta r \sim \delta t \geqslant r^{1/2}$ can be
easily obtained. It can also be gained by directly comparing the
energy formula $E\sim r^{3}T^{4}\sim T\left(r^{3}T^{3}\right)$ for
the photon gas system with Eq.\eqref{1}. By contrast, to count the
holographic degrees of freedom or entropy that is applicable to
\textquotedblleft black hole computers\textquotedblright
\cite{seth4}, we have $\delta r \sim r^{1/3}$ and thus $\delta t\sim
r$. The corresponding UV cutoffs of these systems are obtained from
$\Lambda=(\delta r)^{-1}$.
\[ \begin{CD}
 \text{UV cutoff $\Lambda$: } 0 @>\text{LQFT}>S\leqslant A^{3/4}> \left( rl_p\right)^{-1/2}
 @>\text{New physics?}>S\leqslant A> \left( rl_p^2\right)^{-1/3}
 \end{CD} \]
As argued by Cohen et al, the physics with energy below the UV
cutoff $\Lambda\leqslant r^{-1/2}$ is well-described by LQFT. It was
also pointed out by Hsu \cite{hsu} that when $\Lambda>r^{-1/2}$, the
gravitational corrections to the energy of a LQFT system will be too
large and lead it to undergo gravitational collapse, which makes a
LQFT description invalid. The physics beyond LQFT from $\Lambda\sim
r^{-1/2}$ to $\Lambda\sim r^{-1/3}$ is still obscure now. It might
be some new physics constituting a necessary part of quantum
gravity.

Note that the entropy bound $A^{3/4}$ is for conventional matter
configurations, while $A$ is for the back holes of the same energy
and is considered as the maximum entropy contained in the region in
the spirit of holographic principle. Consider a system or a star
composed of conventional matter with lesser entropy undergoes
gravitational collapse to form a black hole with the area entropy
$A$. It involves a drastic change in its interior metric from
near-flat to an extreme one. What happens in such a collapse process
in the context of quantum gravity? And what could fill such an
entropy gap between $A^{3/4}$ and $A$? They are both interesting but
difficult questions to be further explored and are beyond the scope
of this Letter. It is worth to note that the so-called
\textquotedblleft monster configurations\textquotedblright\ which
originated from a curved space consideration seem to be a candidate
for filling up the gap between $A^{3/4}$ and $A$. See the original
works of Hsu et al \cite{hsu3,hsu4} for details.

\section{Examples\label{s3}}
\subsection{Universe as a supercomputer\label{s3.1}}
Lloyd has investigated the universe from a quantum computational
viewpoint in \cite{seth1,seth2}. It was found that the universe has
performed about $10^{120}$ \textquotedblleft ops\textquotedblright\
or \textquotedblleft ticks and clicks\textquotedblright\ since the
big bang. In this section, we shall explain and clarify this idea
using the space-time uncertainty relation \eqref{2}.

Since a computer has $\frac{r^{3}}{\left(\delta r\right)^{3}}$
working bits with each responding $\frac{t}{\delta t}$ times within
a time interval $t$. Thus the total number of operations that can be
performed in a supercomputer of radius $r$ over time $t$, in other
words, the number of events that can occur in this volume of
space-time is
\begin{align}
\sharp\equiv\frac{r^{3}}{(\delta r)^{3}}\frac{t}{(\delta
t)}\leqslant\frac {rt}{\pi}. \label{3}
\end{align}
The universe, like any computer produced in human factories, is
obeying GR and QM and should be restricted by the above
quantum-gravitational limit. It is a rather huge number when
completed as $\frac{rt}{\pi l_{p}^{2}}$ and will be able to support
any miracle that has happened in the history of the universe. This
computational limit was obtained by Lloyd \cite{seth3} from
$\sharp=\frac{t}{ \pi/(2E_{total})}\leqslant\frac{rt}{\pi}$. The
equivalence of our derivation to this one is easily proved, by
noticing that $E_{total}=\varepsilon\frac{r^{3}}{\left( \delta
r\right)^{3}}$. The derivation in \cite{seth3} is directly based on
the energy limitation from GR. By contrast, our derivation reveals
more subtleties on this issue, making one compare easily the
differences between LQFT systems and holographic systems in
computational aspect. That is, LQFT systems have at most $A^{3/4}$
computational units with each processing at a rate $r^{-1/2}$, while
holographic objects like black holes have $A$ units processing at
the rate $r^{-1}$. Though these systems as whole are subject to the
same bound on the information processing rate, which is
$A^{3/4}r^{-1/2}\sim Ar^{-1}\sim r$ as found in \cite{hsu2}, a
single quantum bit of a LQFT system runs more efficiently than these
in holographic systems.

Now consider a spatial-flat universe which is homogeneous and
isotropic and is described by the FRW metric. The corresponding
Friedmann equations read
\begin{align}
3\left( \frac{\dot{a}}{a}\right) ^{2}=8\pi \rho,  \label{4} \\
\dot{\rho}+ 3\left( 1+w\right)\frac{\dot{a}}{a}\rho=0. \label{5}
\end{align}
Here $w$ is defined by the effective equation of state (EoS) of the
constituents of the universe, $p=w\rho$.

The first Friedmann equation \eqref{4} can be written as $\rho
=\frac{3}{8\pi }r_{a}^{-2}$, where $r_{a}$ is the radius of the
dynamical apparent horizon of the universe. It is equal to the
Hubble radius $r_{h}\equiv(\frac{\dot{a}}{a})^{-1}$ in the flat
universe. Without confusion we shall write them both as $r$. The
total energy confined within the apparent horizon is thus
$\rho(\frac{4\pi}{3}r^{3})=\frac{r}{2}$, as the same amount as the
critical energy to form a black hole of the same size. Since the
apparent horizon is argued to be a causal horizon
\cite{caironggen2}, it is natural to view the FRW universe as a
supercomputer of radius $r$ which is running at its ceiling running
speed determined by GR and QM. It is easy to compute out the number
of \textquotedblleft ops\textquotedblright\ performed by the
universe since the big bang, through the formula
$\sharp=\frac{1}{\pi l_p^2}\int r(t)dt$. From the Friedmann
equations, the energy used to execute computations is just these
spreading in the universe: radiation, dusts, dark energy, or others.
Thus we know the universe keeps doing computations and what it
computes is just the dynamical evolution of itself and its
constituents \cite{seth2}. Only little amount of its energy is
employed by human beings to perform digital computations. For more
details of the computational aspects of the universe, see
\cite{seth2,jack2}. A point worthy of remark is that the EoS of the
constituent greatly affects the calculation of the \textquotedblleft
ops\textquotedblright\ numbers executed by the universe, since it
determines the behavior of $r(t)$ by virtue of $\dot{r}\left(
t\right) =\frac{3}{2}\left( 1+w\right)$ which can be derived from
the Friedman equations.

Knowing $E_{total}=\frac{r}{2}$ for the universe, the space-time
uncertainty relation is saturated as
\begin{align}
\left(\delta t\right)\left(\delta r\right)^{3}=\pi r^{2}.\label{6}
\end{align}
It characterizes the quantum fluctuations of the constituents
uniformly distributed in the universe. We generally write $\delta
r=cr^{\alpha} $, $\delta t=c^{-3}r^{2-3\alpha}$ obeying the
relation, where $c$ is a parameter of order $1$. Different
constituents of the universe such as local quantum fields or some
unknown holographic contents shall involve different type of
space-time fluctuations, with the choice of $\alpha=\frac{1}{2}$ for
radiation and $\alpha=\frac{1}{3}$ for holographic constituents
\cite{jack1}.

To the universe as a supercomputer, the size of any functional cell
must be smaller than the whole size $r$ of the supercomputer, thus
there is $\delta r\leqslant r$. Meanwhile, QM requires the lowest
definable energy for an independent degree of freedom of the system
confined in a box is the IR cutoff energy $r^{-1}$, thus
$\varepsilon\geqslant r^{-1}$. The two limitations lead to the upper
and lower bound on the space-time uncertainties. By applying
Eq.\eqref{6}, we determine the range of the parameter as $\alpha \in
\lbrack \frac{1}{3},1]$. The total information storage capacity
$\frac{r^{3}}{\left( \delta r\right)^{3}}\sim r^{3-3\alpha}$ of the
computer can only range from $1$ to $A$, as expected by the
holographic principle. Furthermore, Eq.\eqref{6} directly reveals a
space-time noncommutative property usually emerging in
quantum-gravitational models \cite{MST}. An evidence is that the
spatial size $\delta r$ of an independent operational unit of the
system increases with its energy $\varepsilon\sim (\delta t)^{-1}$,
which is a typical sign of the space-time noncommutativity.

\subsection{Thermodynamics properties of the universe\label{s3.2}}
In this section, we show that when employing the space-time relation
\eqref{6}, some of the thermodynamics properties of the universe can
be extracted directly. Barrow \cite{barrow} has investigated in
detail the cosmological counterpart of the entropy gap between
matter and radiation entropies and the Bekenstein-Hawking entropy.
We implement such a cosmological entropy gap in our consideration.
Both the radiation-dominated universe with information storage
capacity $A^{3/4}$ and the holographic universe are referred to.
More interestingly, when limited to the holographic case, the
deduced expressions have a similar form with these in
\cite{pad1,pad2}, which aimed to explore the profound physical
connections between thermodynamics and gravity.

Consider a spatial-flat FRW universe filled with a perfect fluid
with EoS $p=w\rho$. The energy-momentum tensor of the fluid is of
the form
\begin{align}
T_{ab}=\left( \rho+p\right) u_{a}u_{b}+pg_{ab}, \label{7}
\end{align}
where $u_{a}$ satisfies $u_{a}u^{a}=-1$. The energy $U$ is the
integral of the energy density over the volume enveloped by the
apparent horizon
\begin{align}
U\equiv\int T_{ab}u^{a}u^{b}dV=\int \rho dV. \label{8}
\end{align}
From the Friedmann equation \eqref{4}, we find $U=\frac{r}{2}$ and
$dU=\frac{1}{2}dr=\frac{1}{3}\rho dV$. This relation is different
from the conventional one $dU=\rho dV$, due to the fact that the
energy here is not proportional to the 3-dimensional volume of the
system, but to its radius. On the other hand, the entropy of the
universe is evaluated as
\begin{align}
S=r^{3}\left( \delta r\right)^{-3}=c^{-3}r^{3-3\alpha}. \label{9}
\end{align}
Having energy $U$ and entropy $S$, the temperature can be obtained
from the thermodynamical law $dU=TdS-pdV$, that is \footnote{Here
the temperature can be negative when $w<-1/3$ leads to an
accelerating universe. One who is not accustomed to a negative
temperature can define the temperature to be positive like that in
\cite{pad1}, i.e., $T\equiv\left\vert \beta\right\vert^{-1}$, where
$\beta F=\beta U-S$. For implications of $\beta\gtrless0$, see
\cite{pad1}. Here it is simply an indication of the acceration or
deceleration of the universe expansion.}
\begin{align}
T=\frac{dU+pdV}{dS}=\frac{1}{3-3\alpha}\left( 1+3w\right)
\frac{1}{2} c^{3}r^{-2+3\alpha}.  \label{10}
\end{align}

Combining Eq.\eqref{9} and Eq.\eqref{10}, we find
$TS=\frac{1+3w}{3-3\alpha}U$. Write it in a more general form
\begin{align}
\begin{split}
TS &=\frac{1}{3-3\alpha}\int(\rho+3p)dV \\
&=\frac{2}{3-3\alpha}\int\left(
T_{ab}-\frac{1}{2}T_{\,c}^{c}g_{ab}\right) u^{a}u^{b}dV. \label{11}
\end{split}
\end{align}
Then the free energy of the system is given by
\begin{align}
F =U-TS =\frac{1}{3-3\alpha}\int[\frac{1}{8\pi}R+\left(
1-3\alpha\right) T_{ab}u^{a}u^{b}]dV.  \label{12}
\end{align}
Here the Einstein equation has been used.

Now we consider a radiation-dominate universe. Radiation has the EoS
$w=\frac{1}{3}$. Together with its characteristic parameter
$\alpha=\frac{1}{2}$ and the formula above Eq.\eqref{11}, we obtain
$TS=\frac{1+3w}{3-3\alpha}U=\frac{4}{3}U$. So the free energy is
$F=U-TS=-\frac{1}{3}U$, exactly consistent with the thermodynamical
property of photon gas. This observation is interesting, since it
reveals that the application of $\alpha=\frac{1}{2}$ is essential
for radiation. Surely, $T\sim r^{-2+3\alpha }\sim r^{-1/2}$
coincides with a familiar temperature-time relation for the
radiation-dominated universe in standard cosmology. \footnote{For a
radiation-dominated universe, there is $r=2t$ and $\rho\sim a^{-4}$.
Thus the temperature is evaluated as $T\sim
a\left(t\right)^{-1}\sim\rho^{1/4}\sim r^{-1/2}\sim t^{-1/2}$.} In
addition, having the temperature $T\sim r^{-1/2}$, the
thermodynamics law $dU=TdS-pdV$ immediately requires the entropy
should be $A^{3/4}$ to make sure that $TdS$ is comparable to
$dU=\frac{1}{2}dr$. This gives another definite illustration that
the information storage capacity of LQFT is $A^{3/4}$ other than in
a holographic form. Obviously the entropy contained within the
apparent horizon increases with the cosmic expansion.

For a holographic universe with $\alpha=\frac{1}{3}$, the
temperature and entropy are respectively $T\sim r^{-2+3\alpha }\sim
r^{-1}$ and $S\sim r^{3-3\alpha }\sim A$, the same as the
thermodynamics for the de Sitter universe. Moreover, in this
holographic case the $\left( 1-3\alpha\right)$ term in Eq.\eqref{12}
vanishes, thus the free energy becomes an integral of the scalar
curvature. We notice that for this case our formulae \eqref{8},
\eqref{11} and \eqref{12} have similar forms with these in
\cite{pad1,pad2}, which asserted that there is a close relationship
between thermodynamical variables and geometrical variables, such as
entropy density and gravitational acceleration \footnote{For
gravitational systems we can not only define the internal energy
$U\equiv\int T_{ab}u^{a}u^{b}dV$, but also the energy source of
gravitational acceleration $E\equiv \int\left(
T_{ab}-\frac{1}{2}T_{\,c}^{c}g_{ab}\right) u^{a}u^{b}dV$, thus a
close relation $E=\left(3-3\alpha\right)TS$ between $TS$ and $E$ can
be observed here. Recall that for a Schwarzschild black hole, we
have the thermodynamics description: $dM=TdS$, where
$M=\frac{r}{2}$, $T=\frac{1}{8\pi M}$, $S=4\pi M^{2}$. This gives
$2TS=M=E$ like that in \cite{pad1}, corresponding to the holographic
case where $\alpha=\frac{1}{3}$. }, free energy density and scalar
curvature. Inspired by the black hole thermodynamics
\cite{hawking,bekenstein}, many works have been devoted to explore
more profound physical connections between thermodynamics and
gravity, and to associate the notions of temperature and entropy
with the spacetimes having horizons
\cite{jacobson,caironggen2,pad1,pad2}. Whatever, the discussions in
\cite{pad1,pad2} are mainly around static spacetimes and based on an
ansatz for gravitational entropy. So we expect these expressions of
thermodynamical variables here and in \cite{pad1,pad2} should imply
some general properties of the connections between thermodynamics
and gravity.

\subsection{Holographic dark energy \label{s3.3}}
The quantum fluctuations in our scenario have a close relation with
previous HDE models \cite{limiao,nojiri,hsu,m2,caironggen1,wei}.
Both the LQFT type and holographic type of quantum fluctuations have
shown up in the derivation of the energy density behavior
$\rho_{\Lambda}\sim r^{-2}$ which is essential for these models. The
derivations take the similar forms as
\begin{align}
\rho_{\Lambda}&\sim\frac{\varepsilon}{\left(\delta
r\right)^{3}}\sim\left(
r^{-1/2}\right)^{4}\sim r^{-2},  \label{13} \\
\rho_{\Lambda}&\sim\frac{\varepsilon}{(\delta r)^{3}}\sim\frac{r^{-1}}{(r^{1/3})^{3}}%
\sim r^{-2}.  \label{14}
\end{align}
It has led to puzzles there why the same density behavior arises
from different derivations \cite{caironggen1,wei}. Here we point out
the differences are superficial. Since the maximum realizable energy
of a system is always the critical energy to form a black hole
\footnote{To introduce it as a choice of dark energy, one often
conjectures that the total quantum zero-point energy in a region
should not succeed the energy of a black hole of the same size
\cite{limiao}.}, one always has $\rho_{\Lambda}\sim r/r^{3}\sim
r^{-2}$, despite where one starts from. In other words, we can
generally compute the energy density associated with the space-time
uncertainty relation as
\begin{align}
\rho_{\Lambda}\sim\frac{\varepsilon}{\left( \delta r\right)
^{3}}\sim\frac{1}{\left( \delta t\right) \left( \delta r\right)
^{3}}\sim r^{-2}.  \label{15}
\end{align}
Obviously, according to Eq.\eqref{6}, the derivation to this type of
energy density is independent of certain choices of $\delta r$ and
$\delta t$.

One may find $\rho_{\Lambda}$ is of the same order of the totoal
energy density $\rho =\frac{3}{8\pi }r^{-2}$ of the universe, and
thus think it may account for the density of dark energy. Actually,
directly taking such an $\rho_{\Lambda}$ as dark energy will lead to
problems. Since $\rho_{\Lambda}/\rho$ is a constant, due to $d\ln
\frac{\rho _{\Lambda }}{\rho }=-3\left( w_{\Lambda }-w\right) d\ln
a$, the EoS of the constituent $\Lambda$ will always trace the
behavior of the effective EoS of the entire universe. As pointed out
by Hsu \cite{hsu}, there is $w_{\Lambda}=0$ in a matter-dominated
universe, thus $\Lambda$ cannot lead to an accelerating universe. To
solve this problem and get a dark energy model having
$w_{\Lambda}<-\frac{1}{3}$, afterwards infrared cutoffs other than
the size of apparent horizon are widely suggested, such as the size
of future event horizon \cite{limiao}, the age of the universe
\cite{caironggen1,wei} or even a mixture of them \cite{nojiri}.

For the rest, though we have shown the derivation of $\rho_{\Lambda
}\sim r^{-2}$ is independent of certain type of quantum
fluctuations, it should be pointed out that Ng \cite{jack1,jack2}
suggested it should be the holographic type of fluctuations $\delta
r\sim r^{1/3}$ responsible for the unconventional dark
energy/matter, because it is different from these of conventional
matter and radiation. This kind of dark energy can be called
\textquotedblleft holographic\textquotedblright, for that it is
related to the holographic entropy. Ng has considered the data from
Hubble Space Telescope to test the existence of such an
unconventional holographic space-time fluctuation.

\section{Conclusion and implications\label{s4}}
By combining GR and QM, the space-time uncertainty relation $(\delta
t)(\delta r)^{3}\geqslant\pi r^{2}$ has been derived from a quantum
computational perspective. The case of $\delta r\geqslant r^{1/2}$
describes the distant fluctuations related to the well-established
LQFT. By contrast, the case of $\delta r\geqslant r^{1/3}$ leads to
the holographic entropy and is attached with some unknown
microscopic physics of gravitational systems. Thought it could be
introduced by various approaches, and Ng has tried to suggest it to
account for the unconventional dark energy/matter, this type of
fluctuations has not yet been understood at a deep lever. The
holographic entropy is a special property attached with the
space-times having horizons. And we also have shown its induced free
energy is relevant to the integral of scalar curvature without
matter term present, like that in \cite{pad1} which aimed to explore
the possible thermodynamical properties of gravity. Does $\delta
r\geqslant r^{1/3}$ really characterize the quantum fluctuations of
gravitons or say space-time itself? The question is surely worthy of
further study.

We give several examples where the space-time uncertainty can be
used, including the information storage and computational capacity
of the universe, the thermodynamics properties of the universe, and
the origin of holographic dark energy. Each topic is of interest in
itself. Here we don't intend to discuss the details of them, but
only focus on extracting typical characteristics of these topics by
virtue of the space-time uncertainty relation and comparing them
with the known results there. The universality and validity of the
proposed relation are thus exemplified.

\section*{acknowledgments}

We would like to thank J. L. Li for useful discussions. The work is
supported in part by the NNSF of China Grant No. 90503009, No.
10775116, and 973 Program Grant No. 2005CB724508.

\end{document}